\documentclass{PoS}
\newcommand{\be}{\begin{equation}}
\newcommand{\ee}{\end{equation}}
\newcommand{\beq}{\begin{eqnarray}}
\newcommand{\eeq}{\end{eqnarray}}
\usepackage{graphicx}
\usepackage{epsfig}
\usepackage{epsf}

\title{A novel method for the determination of hadron excited states in Lattice QCD applied to the nucleon}

\ShortTitle{A novel fitting scheme - Nucleon excited states}
\author{\speaker{Constantia Alexandrou}\\
  Department of Physics, University of Cyprus, P.O. Box 20537,
 1678 Nicosia, Cyprus\\
        E-mail: \email{alexand@ucy.ac.cy}}
\author{C. N. Papanicolas\\
        Department of Physics and Institute of
  Accelerating Systems \& Applications, University of Athens, Athens, Greece\\
and\\ The Cyprus Institute, 15 Kypranoros St., 1645 Nicosia, Cyprus \\ 
       E-mail: \email{cnp@iasa.gr, cnp@cyi.ac.cy}}
\author{E. Stiliaris\\
        Department of Physics and Institute of
  Accelerating Systems \& Applications, University of Athens, Athens, Greece\\ 
       E-mail: \email{stathis@iasa.gr}}
\abstract{A novel method for the  precise
identification and  determination of the energies  that
contribute in the spectral decomposition of lattice correlators is presented.
The method is based on statistical concepts and it relies heavily on simulation techniques. 
 The $\eta_c$ correlator is analyzed within this method and the results 
obtained are 
 compared 
to a previous analysis based on  Bayesian statistics.
 An analysis of the
 nucleon local two-point correlators
leads to the identification of  the excited states in the positive and negative
channels. A discussion on the Roper is included.
}
\FullConference{The XXVI International Symposium on Lattice Field Theory\\
		 July 14-19 2008\\
		 Williamsburg, Virginia, USA}
\begin{document}

\section{Introduction}

A significant component of hadronic physics research focuses  on
the understanding of the excitation spectrum of hadrons and 
in particular that of the proton~\cite{NSTAR01}. 
 On the experimental side, high quality data are becoming available 
 largely due to the advances in
accelerator and instrumentation technologies.
Lattice QCD simulations
using new algorithms and faster computers  yield high-precision
results close to the chiral limit, where
chiral perturbation theory
can be reliably applied thus yielding
physical results that can be  compared
to experiment.
 However,
achieving a high precision necessitates a robust analysis of
simulation data keeping systematic errors under control.
Such an analysis  typically involves a fitting procedure. 
The quantities of interest are
  masses of hadrons, decay constants, form factors
and other hadronic matrix elements. In most cases
one uses a subset of the lattice data and truncates the theory
to a small number of fitting parameters.
For example,  in the case of two-point functions one discards simulation results involving short times.
Such a truncation
eliminates important
 information about excited states encoded in these two-point
functions. 
The goal of the current  study is to apply a method to extract the maximum information from lattice measurements.

Various approaches have been proposed to ameliorate the problems of the
conventional fitting procedure most of which are based on  the 
Maximum Entropy Method~\cite{Nakahara, Allton, Lepage, Morningstar}.
In this work we present a new method that relies solely on 
$\chi^2$-minimization with an unbiased evaluation of errors.
 It has been developed
and applied in the context of the analysis of electroproduction data
in the nucleon resonance region~\cite{Stiliaris} but, as demonstrated here, it is of
general applicability. The method has as a minimal
requirement that the parameters to be determined are linked in an
explicit way to the data. There is no
requirement that this set of parameters provide an orthogonal
basis. Moreover these parameters can be subjected to explicit
constraints. 
The main advantages of the method is that it requires no prior
knowledge (priors) other than the spectral decomposition of lattice
correlators
 nor any assumption on the level of truncation. The
data set alone determines the information that can be extracted.

In order to demonstrate the method we analyze the $\eta_c$-correlator 
to extract the three lowest states~\cite{Davies} and compare
the results of our method  to the values extracted using the method of 
Ref.~\cite{Lepage} on the same data.
Local nucleon correlators on standard lattices
 are easy to produce and are 
readily available.
We show that the method is well suited to extract the excited states of the nucleon in
both the positive and negative parity channels. Using dynamical twisted
mass fermions as well as dynamical Wilson fermions and two different nucleon interpolating fields
we discuss the positive parity excitation of the nucleon and
its connection to the Roper as a function of the pion mass.

\vspace*{-0.3cm}

\section{The method}
\vspace*{-0.3cm}

\begin{figure}[ht]
\vspace*{-0.7cm}\hspace*{-0.4cm}
\begin{minipage}{4.8cm}
\epsfxsize=5.2truecm
\epsfysize=5.5truecm
\ifpdf
{\mbox{\includegraphics[height=5.5cm,width=5.cm]{chi2_distr}}}
\else
{\mbox{\epsfbox{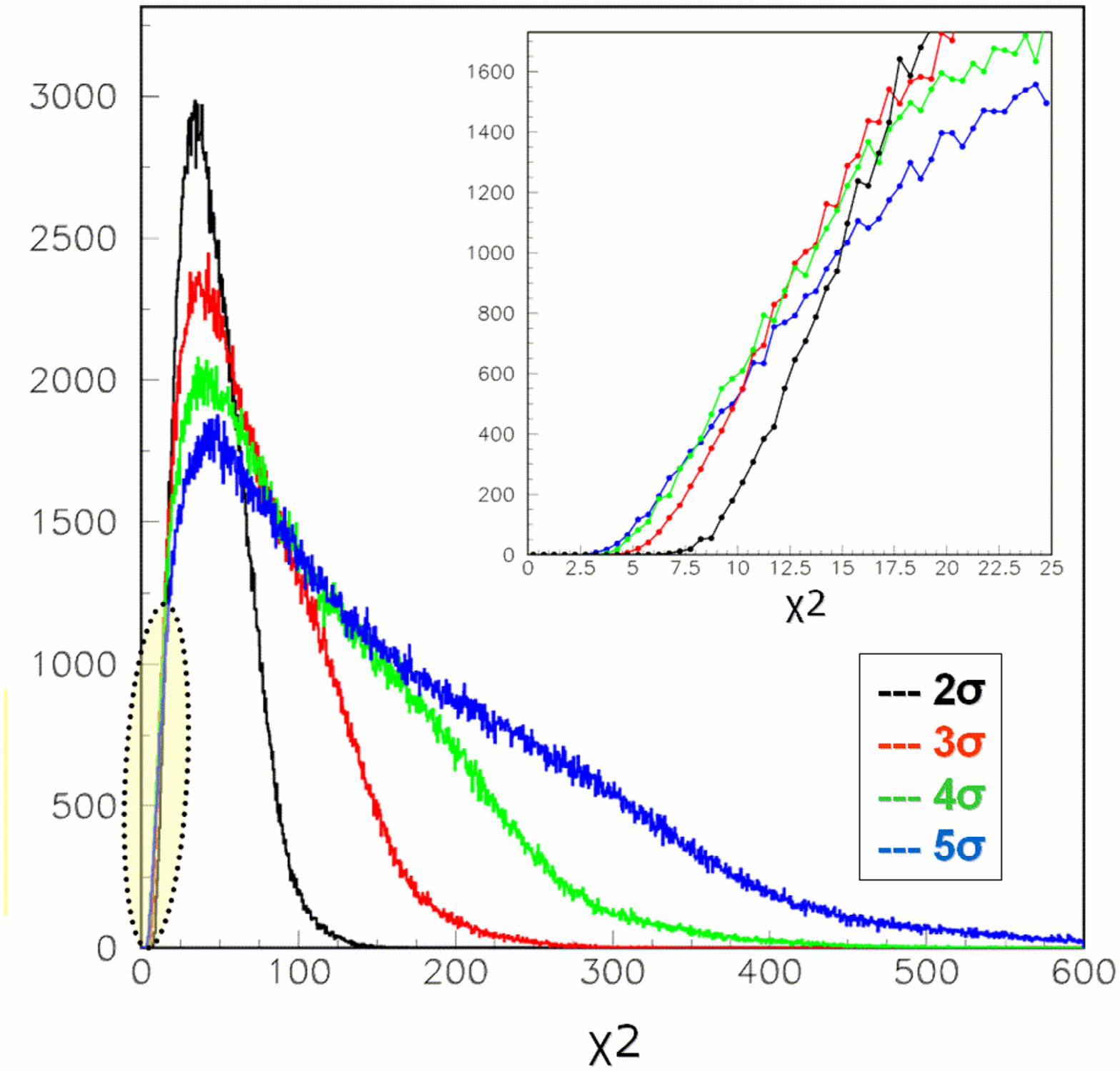}}}
\fi
\caption{The $\chi^2$-distribution as the range of values of the parameter
set $\{B\}$ is increased.}
\label{fig:chi2 distr}
\end{minipage}\hspace*{0.3cm}
\hfill
\begin{minipage}{4.8cm}
\epsfxsize=5.4truecm
\epsfysize=5.5truecm
\ifpdf
{\mbox{\includegraphics[height=6cm,width=5.4cm]{chi2_cuts_m0}}}
\else
{\mbox{\epsfbox{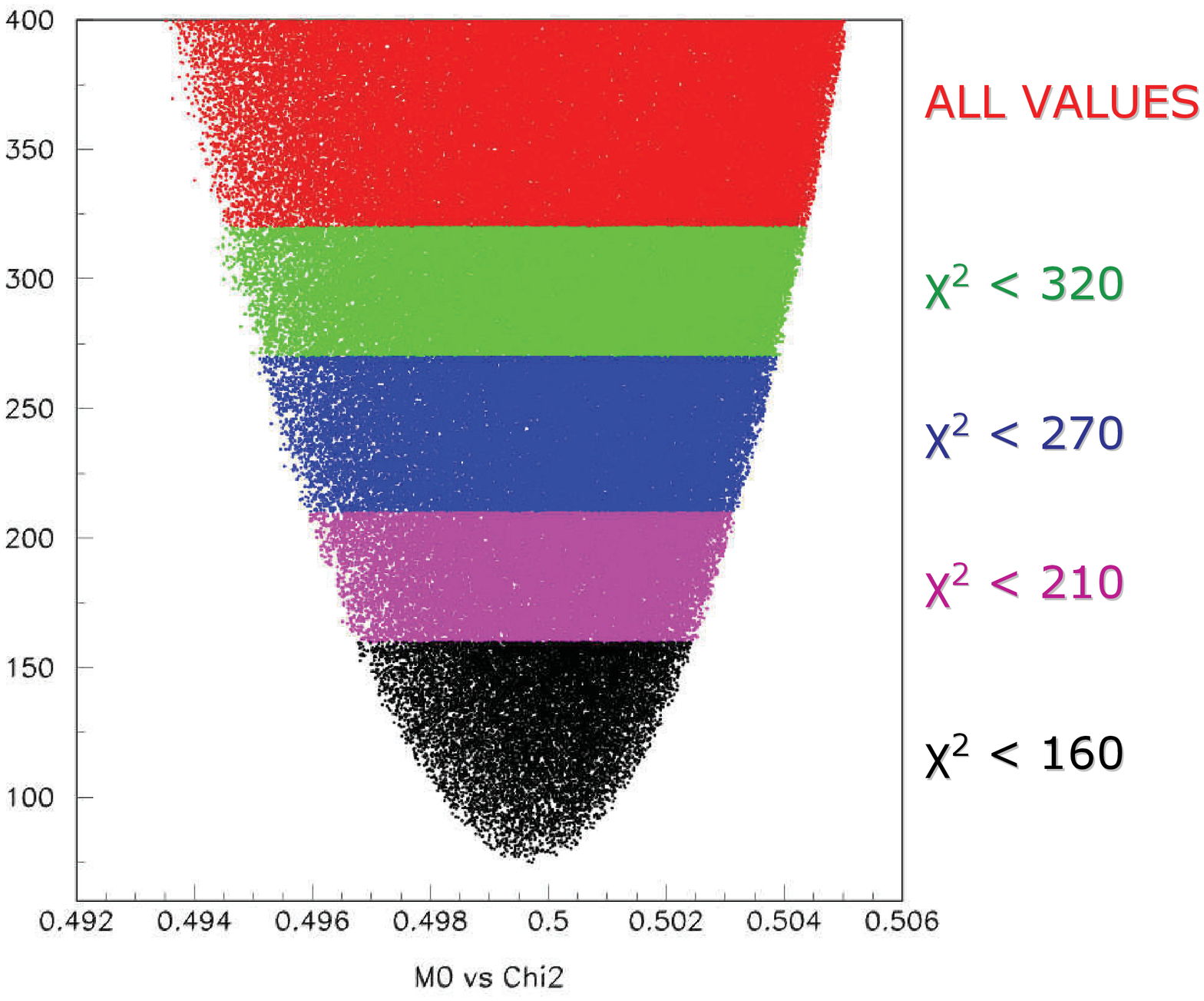}}}
\fi
\caption{The $\chi^2$-distributions for $m_0$ of
a two-point correlator for different $\chi^2$-cuts.}
\label{fig:chi2 cuts}
\end{minipage}\hspace*{0.3cm}
\hfill
\begin{minipage}{4.8cm}\vspace*{-0.4cm}
\epsfxsize=5.4truecm
\epsfysize=5.8truecm
\ifpdf
{\mbox{\includegraphics[height=6.5cm,width=5.4cm]{microcanonical_distr_m0}}}
\else
{\mbox{\epsfbox{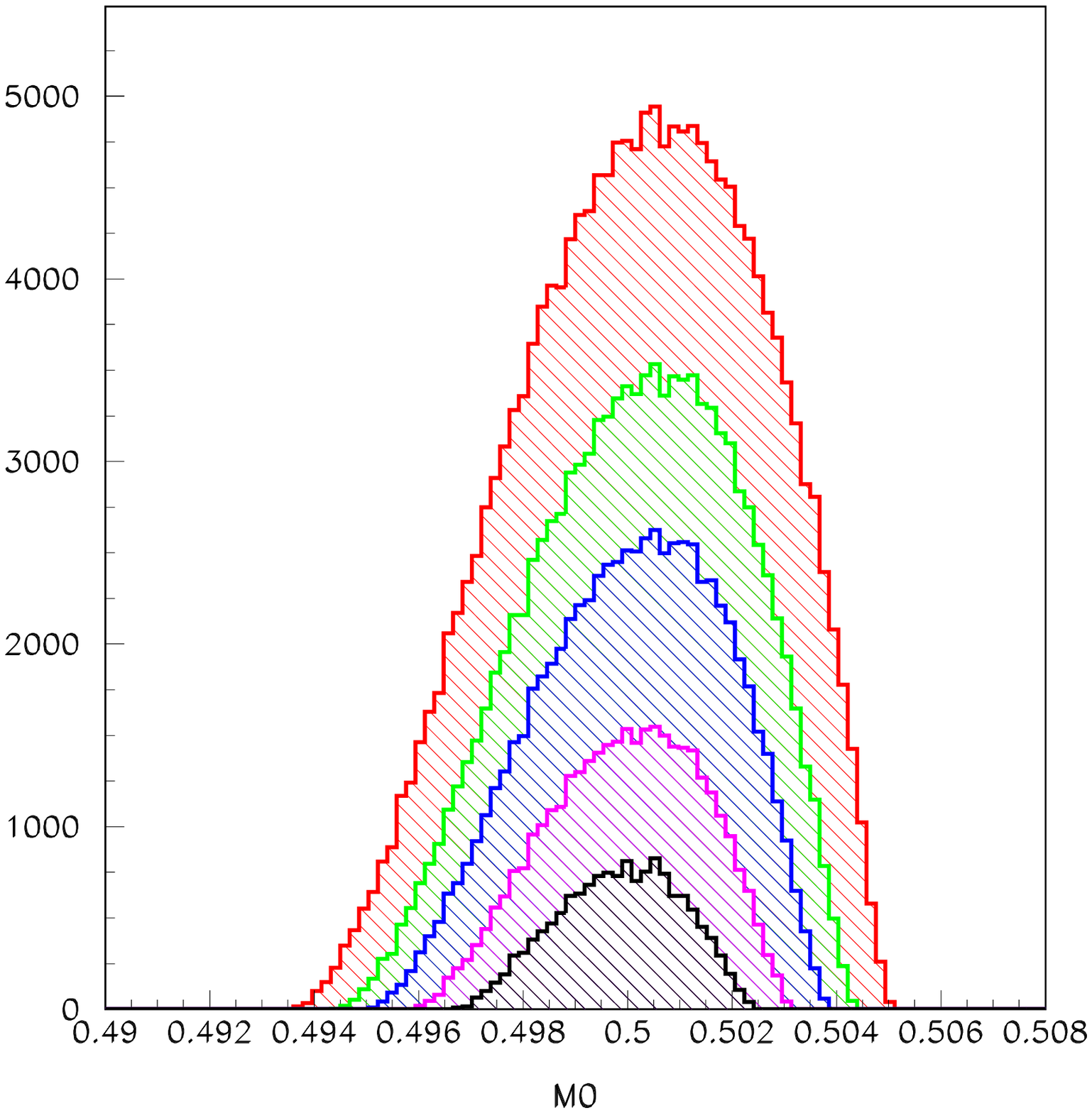}}}
\fi
\caption{The distribution of $m_0$ of a two-point correlator for
different values of  $\chi_{min}$. }
\label{fig:micro distr}
\end{minipage}
%\begin{minipage}{4.5cm}\vspace*{0.3cm}
%\epsfxsize=5.5truecm
%\epsfysize=4.5truecm
%{\mbox{\epsfbox{chi2_cuts_ins.ps}}}
%\caption{The $\chi^2$ distributions as $\chi_{min}$ decreases for a parameter that the
%data are not sensitive on. }
%\label{fig:chi2 cuts ins}
%\end{minipage}
\end{figure}

The method, referred to as ``AMIAS''~\cite{Stiliaris}, 
relies only on the ergodic hypothesis, namely that
any parameter of the theory  can take any possible value 
allowed by the theory and its underlying assumptions. 
The probability of this value representing reality is solely determined by the data.
We assume that all possible values are acceptable solutions, but with varying probability of being true.
If $n$ parameters are needed to describe the data, then each $j$-set  $\{B_1...B_n\}^j$ 
 is
a solution. We assign to each such set a $\chi^2$-value and a probability.\
We then construct an ensemble of solutions and assume that this contains all solutions with finite probability.
The probability distribution for any  parameter assuming a given value is then {\it the solution}.
An ensemble of solutions can be defined as the
collection of solutions, which are characterized by
$\chi^{2}\leq \chi^{2}_{min}+{\cal C} $ with ${\cal C}$ usually taken to
be a constant equal to the effective degrees of freedom of the problem. 
For a solution set $j$ we compute
 $\chi^{2}(n,j)=
     \sum_{k=1}^M\sum_{i}\{\frac{(V_k(t_i) -f(t_i,\{B\}^j)}{w_{k}(t_i)}\}^{2}$,
where $V_k(t_i)$ and $w_k(t_i)$ are $M$ sets of measurements and errors respectively.
For lattice applications this  set of measurements
 is  defined on  temporal
lattice  slices $t_i$
and $f(t_i,\{B\}^j)$ is the spectral decomposition 
for  the relevant correlator
 written in terms of
 the $n$ parameters $\{B\}^j$.
In what follows we further develop the methodology
through the specific problem  of extracting the excited states
from two-point correlators.
The Euclidean time correlator $C(t)$ of an interpolating operator
$J({\bf x},t)$ and its spectral decomposition for zero three-momentum is
\be
C(t)=\sum_{\bf x} <J({\bf x},t) J^\dagger({\bf 0},0)>=\sum_{l=0}^\infty A_l e^{-m_l t}\quad,
\label{correlator}
\ee
and therefore for each $j$-set
 $f(t_i,\{A,m\})=\sum_{l=0}^L A{_l} e^{-m_l t_i}$, where
 $L$ is the highest excited state that the data are
sensitive on and it  is determined by the method.
The exponential dependence is correct if
 we neglect boundary conditions (b.c.). For antiperiodic b.c. a meson
correlator is symmetric about the mid point of the temporal lattice extension
and each exponential is modified to $\cosh m(t-T/2)$.
 We start by assuming a maximum number of $L$ excited states 
in the spectral decomposition of the correlator  
and  select a suitable range of values for each of the
  parameters $A_l\ge 0$ and $m_0<m_1<m_2<\cdots$. 
These ranges define the "phase volume" which the
  Monte Carlo method explores.  
Within the chosen range, we
  uniformly select a value  for each
  parameter. 
 Considering $L$ excited states, i.e. $L+1$
exponentials with $n=2+2L$ parameters, we evaluate the   $\chi^{2}(2+2L,j)$  corresponding to the
  particular solution   using our lattice data.
  We repeat this procedure  a large number  of times, typically a few
  hundred thousand, generating an ensemble of solutions.
The method does not depend on  the choice of 
 $L$, provided that it is sufficiently large. In practice it is chosen so
that the derived results do not change if  instead $L+1$ is used as
a maximum cutoff.
 The computational time required depends critically on the
 choice for the phase volume to be explored.
For the implementation described here  it scales like a power in the number of parameters.
However this can be reduced to about linear if a Markov chain is used to generate
the ensemble.
 It is obvious that the
overwhelming majority of the solutions generated in this fashion
are characterized by very large $\chi^{2}$-values. 
However,  as shown in Fig.~\ref{fig:chi2 distr} a saturation of "good values" is
achieved as the phase volume is enlarged 
that remains unaltered
by exploring a wider range of values for the parameters.

The behavior of a given  parameter $A_l$ or $m_l$ in
the ensemble of solutions can be visualized by plotting
the $\chi^{2}$ versus the value of the parameter. In
 Fig.~\ref{fig:chi2 cuts} we show such a plot for $m_0$, the ground state mass 
of a
two-point correlator. The sensitivity
of the data on this parameter becomes explicit and quantifiable by
applying successive cuts on the $\chi^{2}$- values and
constructing out of the selected population of solutions
histograms, as shown in  Fig.~\ref{fig:micro distr}.
For this parameter on which the data are sensitive on, the  distributions have a well defined maximum independent of the cut 
whereas, as expected, the width depends on the $\chi^2$-cut applied.
%This particular behavior is for a parameter on which the data are sensitive on.
If the data show no sensitivity on a given parameter
then the ensemble of solutions for this particular parameter yields a uniform distribution
independent of the  $\chi^2$-cut.% as shown in Fig.~\ref{fig:chi2 cuts ins}
 Therefore the method does
not treat differently "sensitive" from "insensitive"
exponential terms; they naturally emerge as such. The widths of the
histograms, ranging from very narrow to very wide or infinite,
naturally select and order the various exponentials
according to their sensitivity to the data set.

  \begin{figure}[ht]
%\begin{minipage}{8.5cm}
\epsfxsize=15truecm
\epsfysize=5.5truecm
\ifpdf
{\mbox{\includegraphics[height=10cm,width=15cm]{correlations_mass}}}
\else
{\mbox{\epsfbox{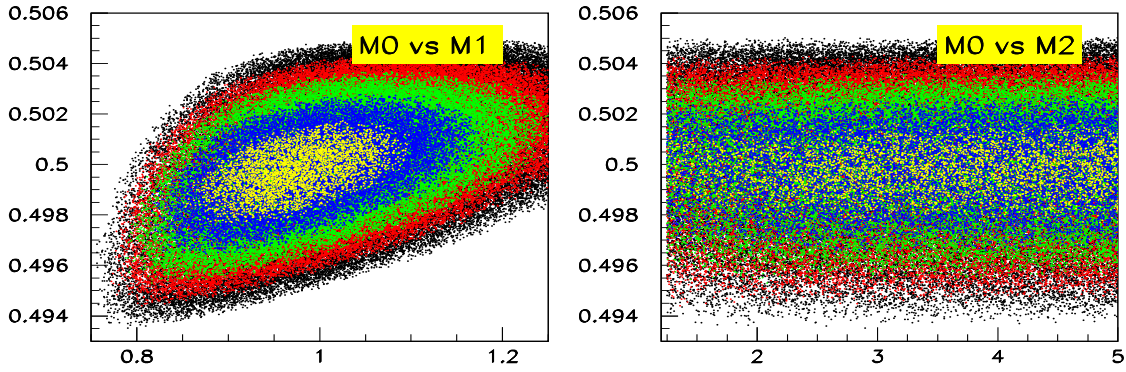}}}
\fi
%\end{minipage}\hspace*{0.5cm}
%\begin{minipage}{8.5cm}
%\epsfxsize=6truecm
%\epsfysize=6truecm
%{\mbox{\epsfbox{corr_onesen_oneinsen.ps}}}
%\end{minipage}
\vspace*{-0.5cm}
\caption{Left: Correlations between two sensitive parameters;
Right: a sensitive and an insensitive parameter.}
%  one sensitive and on insensitive (left).}
\label{fig:correlations}
\vspace*{-0.5cm}
\end{figure}

A central issue that is properly treated in our method, is the handling of
correlations.
All possible correlations are accounted for
by allowing all fit parameters
 to randomly vary and to yield solutions with all allowed values
including the "insensitive" exponential terms. 
The visualization of at least the
dominant correlations is accomplished in a two-dimensional
scatter plot in which the ensemble of solutions is projected on
the plane defined by the values of the parameters and color coded according to
the $\chi^{2}$-value. In
Fig.~\ref{fig:correlations} the strong correlations between $m_0$ and
$m_1$, two of the parameters
on which a two-point correlator depends on, are displayed as compared to the
lack of correlations between $m_0$ and $m_2$
on the latter of which the data do not depend on.

%%%%%%%%%%%%%%%%%%%%%%%
%%  E X A M P L E S  %%
%%%%%%%%%%%%%%%%%%%%%%%
\section{Analysis of $\eta_c$ correlator}

  \begin{figure}[ht]
\vspace*{-0.6cm}
\begin{minipage}{10.cm}\vspace*{-0.3cm}
\epsfxsize=10truecm
\epsfysize=10truecm
\ifpdf
{\mbox{\includegraphics[height=10cm,width=10cm]{etac_distr_new}}}
\else
{\mbox{\epsfbox{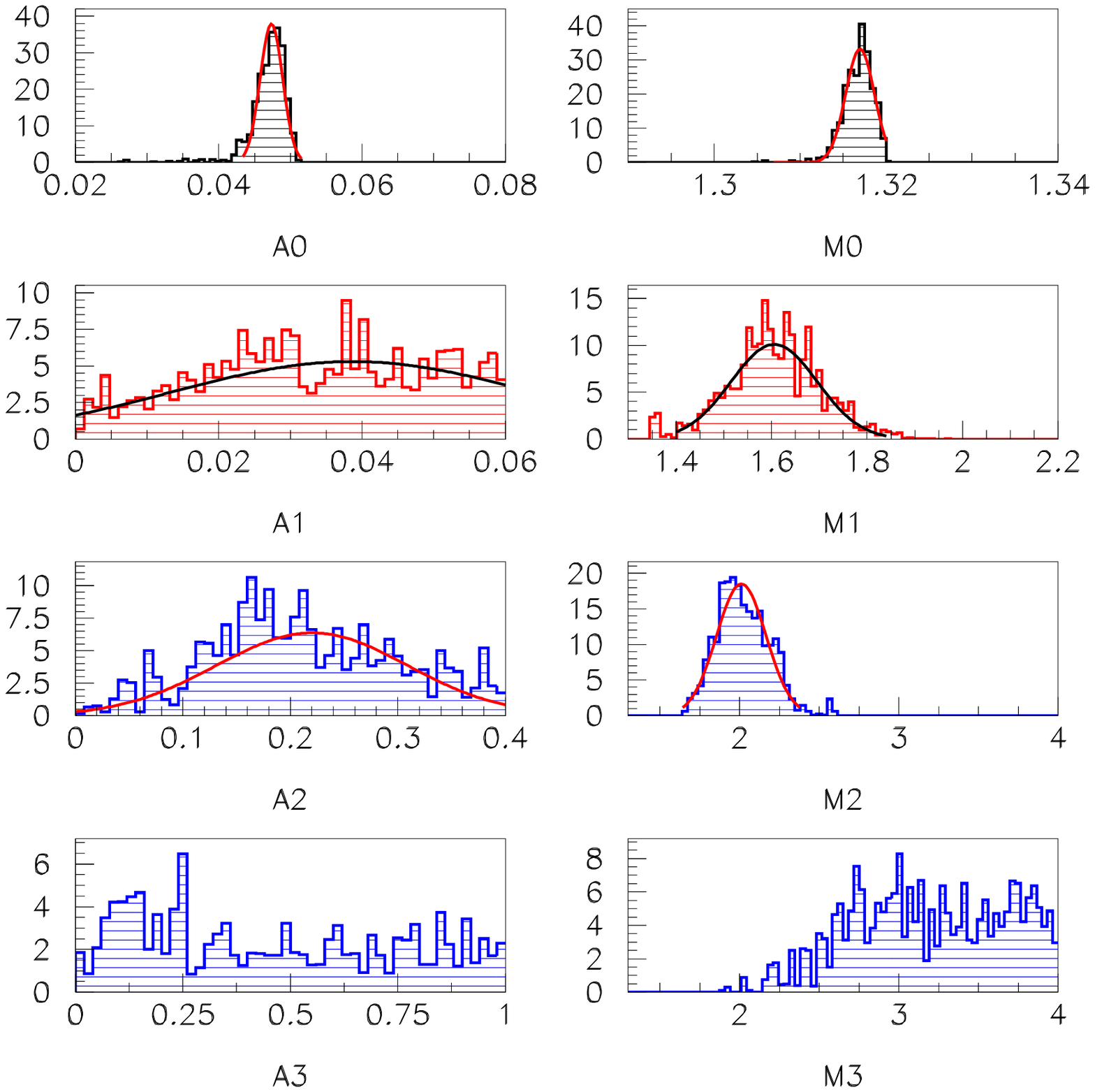}}}
\fi
\vspace*{-1cm}
\caption{The distribution of the amplitudes and masses for $\eta_c$.}
\label{fig:etac}
\end{minipage}
\begin{minipage}{5cm}
\epsfxsize=5truecm
\epsfysize=6truecm
\ifpdf
{\mbox{\includegraphics[height=6cm,width=5cm]{etac_spectrum_new}}}
\else
{\mbox{\epsfbox{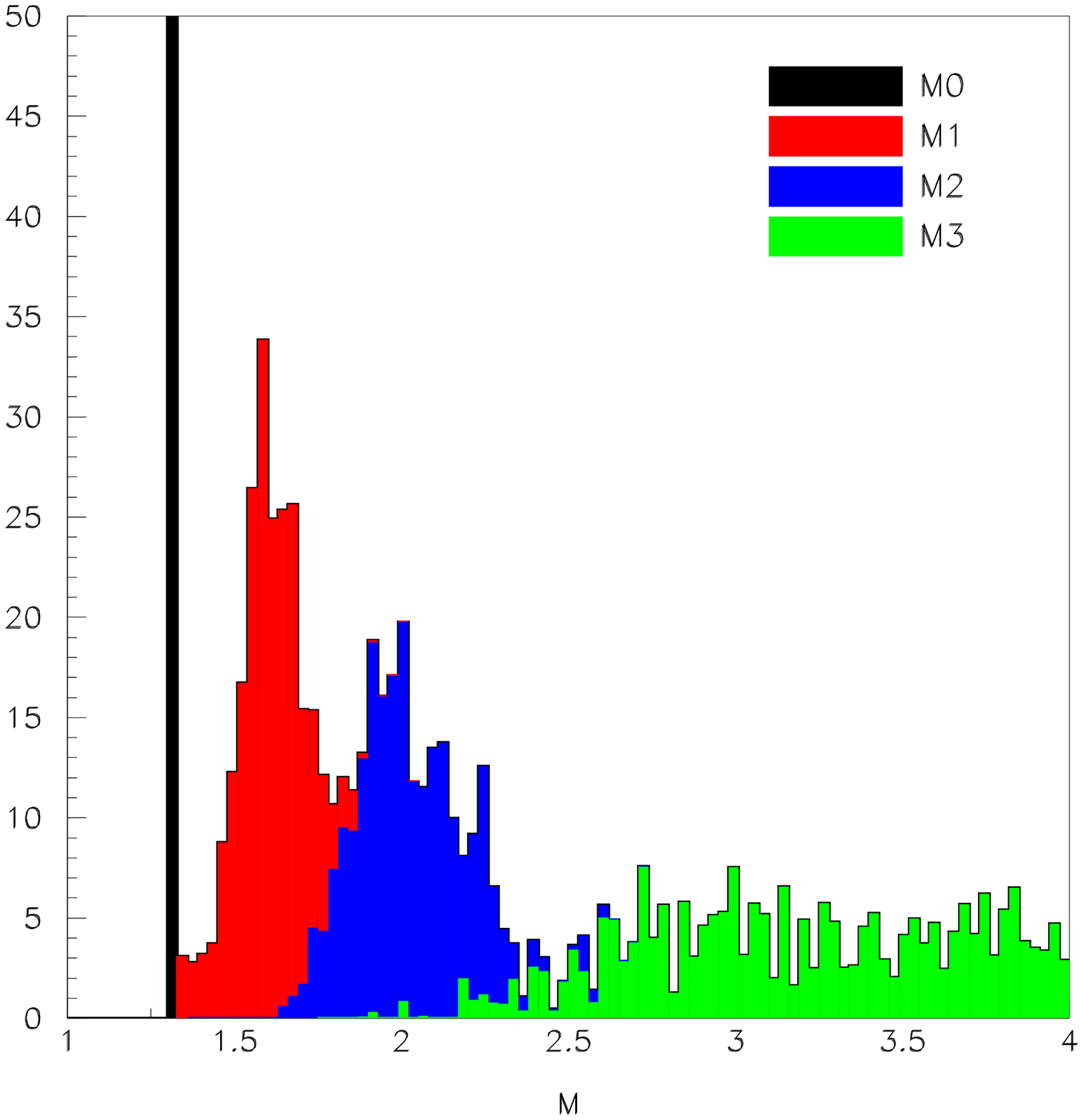}}}
\fi
\caption{The distribution resulting from the ``AMIAS'' ensemble of solutions
yields the mass spectrum of the $\eta_c$ in lattice units. The widths
characterize the uncertainty with which the method determines the parameters.}
\label{fig:etac spectrum}
\end{minipage}
\vspace*{-0.3cm}
\end{figure}

In this section we apply our method to re-analyze the
$\eta_c$-correlator computed on seventy uncorrelated gauge 
configurations on  a lattice of temporal extent $N_t=96$
~\cite{Davies}. These data were analyzed using the method
of Ref.~\cite{Lepage} to extract the masses of the
ground state and the first and second
excited states~\cite{Davies}.
We apply the ``AMIAS'' method using four exponentials i.e. $L=3$. In Fig.~\ref{fig:etac} we show the
distributions for the four amplitudes and four masses. As can
be seen, the three lowest states are determined from the data whereas the
the fourth is undetermined. The values that
we find, starting the fit from the fourth time slice from the source, are:
   1.3171(13), 1.608(9) and 2.010(11) in lattice units,  where the errors are
computed using a jackknife procedure. 
The ``AMIAS''  values are   in agreement with the ones extracted in Ref.~\cite{Davies}
 using the method of
Ref.~\cite{Lepage}, namely   1.3169(1), 1.62(2) and 1.98(22), but
with improved accuracy for the excited states.
Note that, since the mean values are  determined by $\chi_{min}$
and the error using a jackknife procedure, the results are independent of
the parametrization used for the function $f(t_i,\{B\}^j)$. Compatible
errors result from the width of
the ``AMIAS'' distributions~\cite{Stiliaris} as can be seen 
in Figs.~\ref{fig:etac} 
and \ref{fig:etac spectrum}. The latter figure  shows
 the $\eta_c$-spectrum for the 
four states that we fitted.%~\cite{forthcoming}.
 There is a considerable overlap 
of the excited
states and this explains the strong correlations among them.
%The (non-) overlap between the ground and first (second) 
% excited state distributions 
%accounts for the (non-) correlations between $m_0$ and $m_1$ ($m_2$)  shown
%in Fig.~\ref{fig:correlations}.
It is crucial for the determination
of the mean values and the errors to  increase the number of exponentials
beyond the states one is determining.

%\begin{table}[h]
%\begin{minipage}{8.5cm}
%\caption{The  values of $A_n$ and $m_n$ extracted, 
% from the  analysis of the $\eta_c$ correlators compared to the
%values using the method of Ref.~\cite{Lepage}. }
%\label{tab:eta_values}
%\begin{tabular}{|c|c|c|c|c|} \hline
%n & 0 & 1 & 2  & 3  \\
%$m_n$ & 1.3169(1)& 1.62(2)&  1.98(22) & \\
%$m_n$~\cite{Davies} & 1.3150(5)& 1.7047(30)&  1.8166(84)  & \\
%$A_n$ & & &   & \\
 %  \hline
%\end{tabular}
%\end{minipage}
%\end{table}
%\begin{figure}
%\begin{minipage}{8.5cm}
%\epsfxsize=6truecm
%\epsfysize=8truecm
%{\mbox{\epsfbox{etac_spectrum.ps}}}
%\caption{The spectrum of the $\eta_c$ correlator.}
%\label{fig:etac spectrum}
%\end{minipage}
%\end{figure}

\section{Nucleon}
  \begin{figure}[ht]
\vspace*{-0.8cm}
\begin{minipage}{6.cm}
\epsfxsize=6truecm
\epsfysize=6truecm
\ifpdf
{\mbox{\includegraphics[height=6cm,width=6cm]{nucleon_pos_spect}}}
\else
{\mbox{\epsfbox{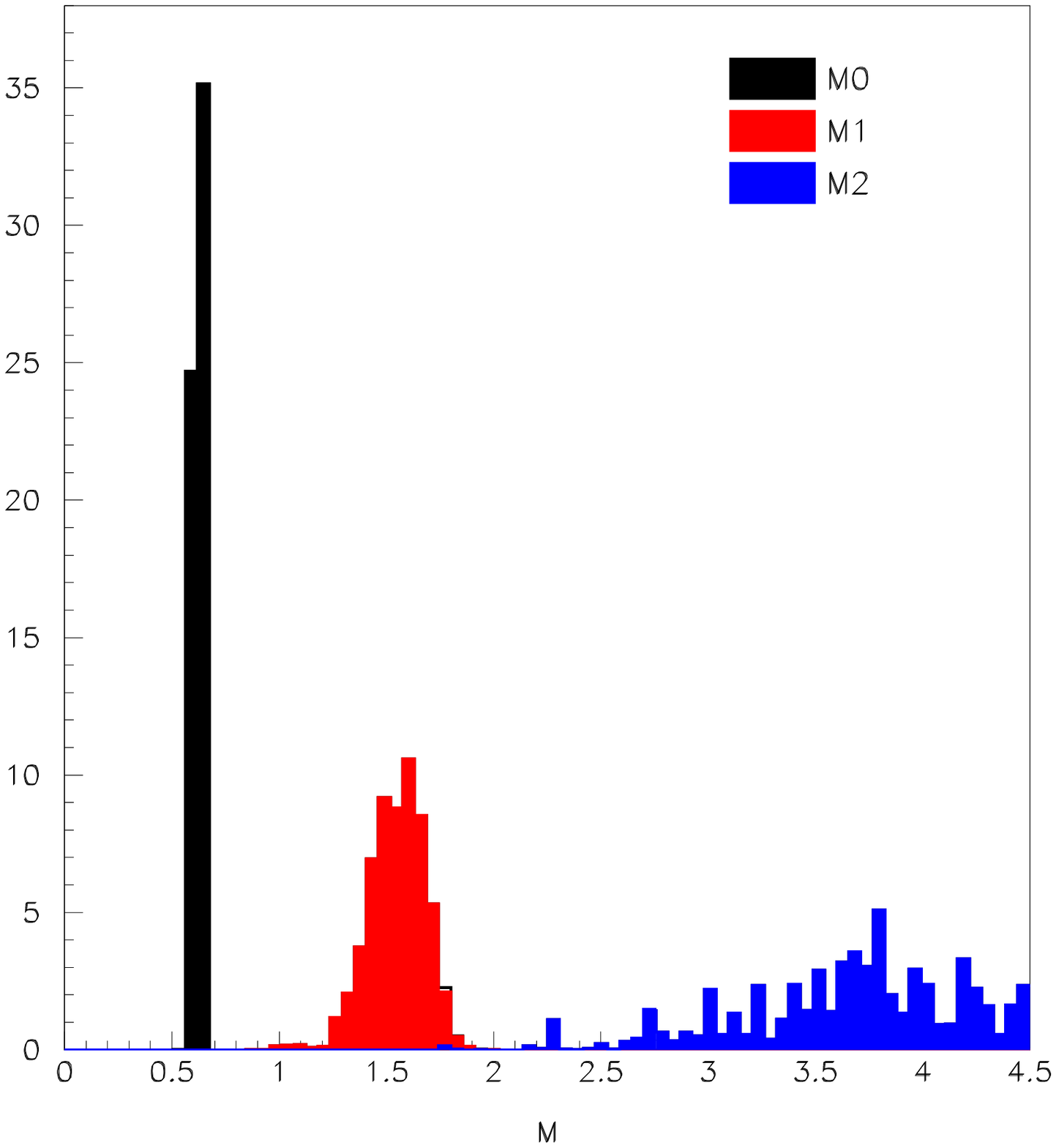}}}
\fi
\end{minipage}\hspace*{2cm}
\begin{minipage}{6cm}
\epsfxsize=6truecm
\epsfysize=6truecm
\ifpdf
{\mbox{\includegraphics[height=6cm,width=6cm]{nucleon_neg_spect}}}
\else
{\mbox{\epsfbox{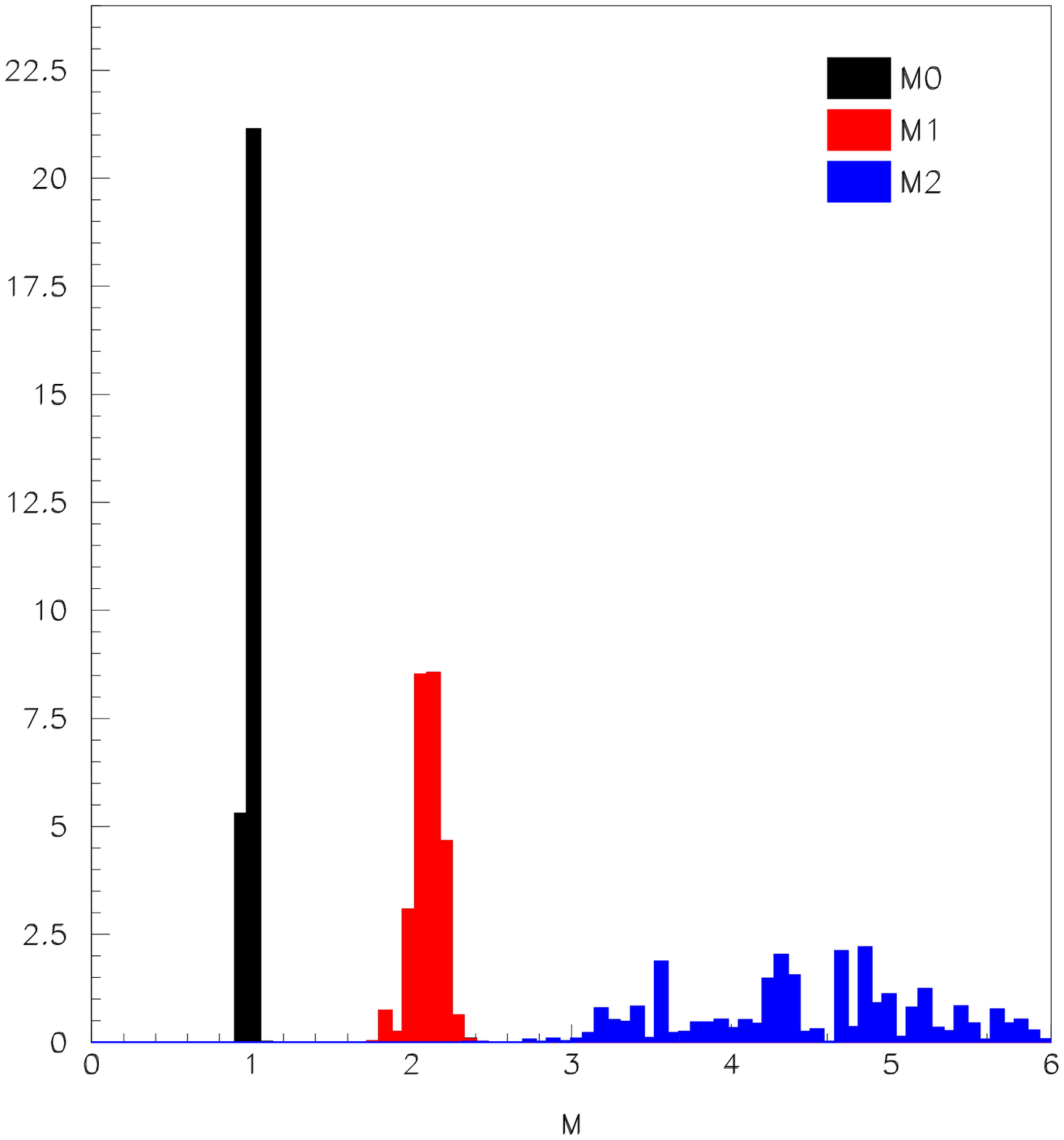}}}
\fi
\end{minipage}
\vspace*{-0.8cm}
\caption{Nucleon mass spectrum in lattice units for the positive (left) and negative (right) parity channels using local correlators with TMF
  at pion mass
484~MeV  on a lattice of spatial length 2.1~fm at $\beta=3.9$.}
\label{fig:nucleon spectrum}
\end{figure}

  \begin{figure}[ht]
\vspace*{-0.3cm}
\begin{minipage}{8.cm}
\epsfxsize=7.5truecm
\epsfysize=7.5truecm
\ifpdf
{\mbox{\includegraphics[height=7.5cm,width=7.5cm]{nucleon_pos}}}
\else
{\mbox{\epsfbox{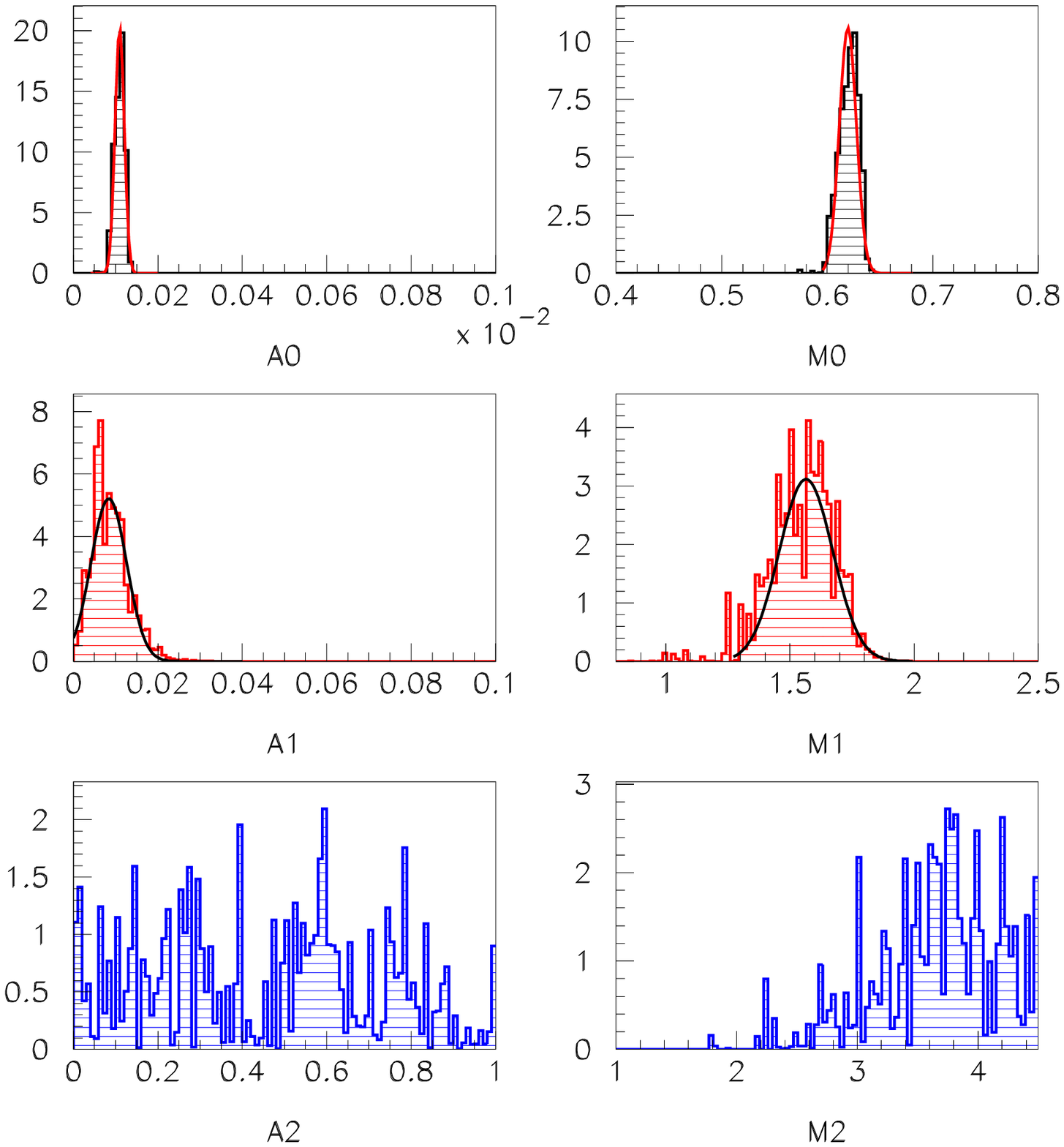}}}
\fi
\end{minipage}
\begin{minipage}{8.cm}
\epsfxsize=7.5truecm
\epsfysize=7.5truecm
\ifpdf
{\mbox{\includegraphics[height=7.5cm,width=7.5cm]{nucleon_avecorr_9_pos}}}
\else
{\mbox{\epsfbox{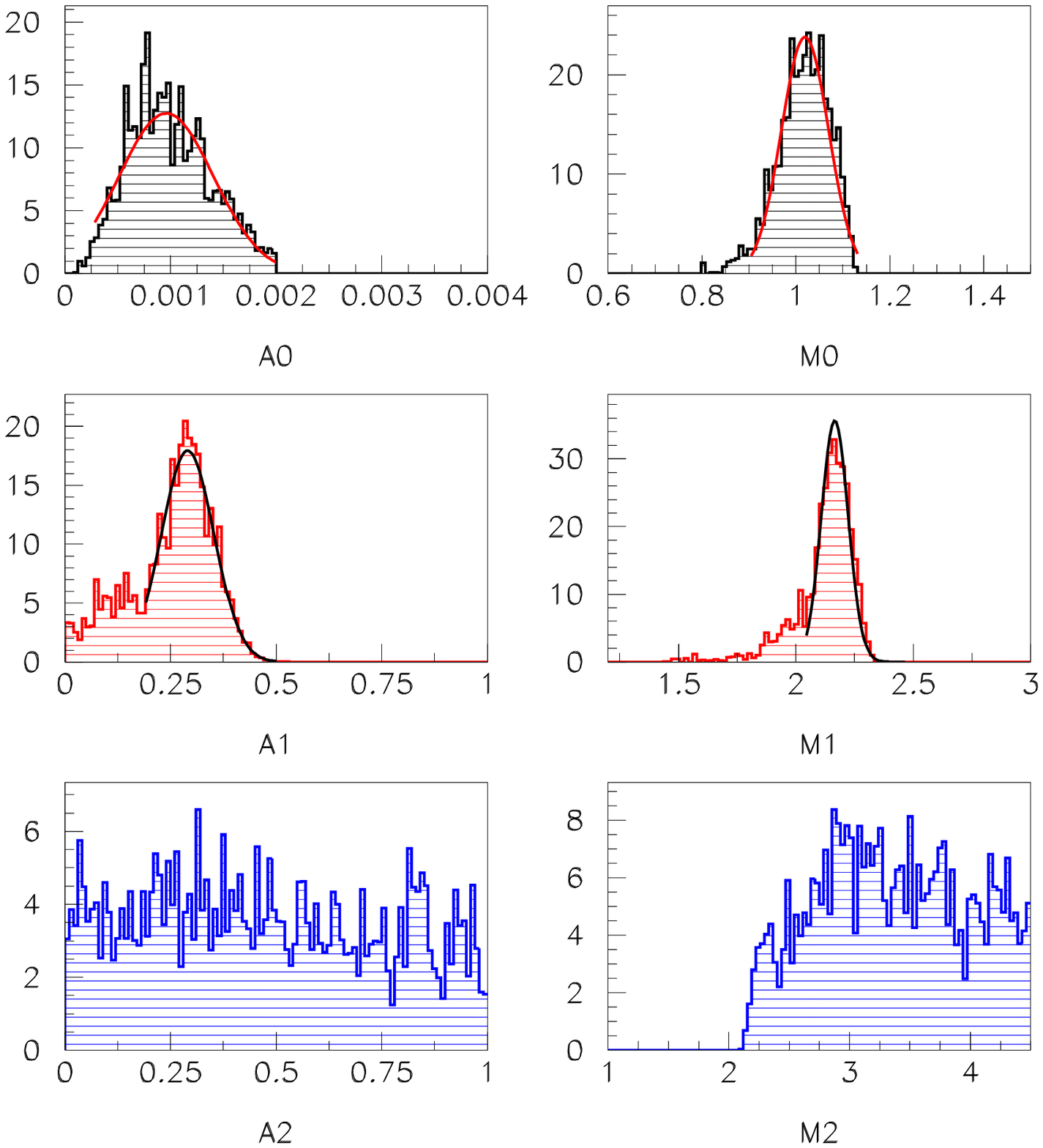}}}
\fi
\end{minipage}
\vspace*{-0.5cm}
\caption{Probability distributions for the amplitudes and masses in lattice
units extracted from
 local correlators computed using $N_F=2$ Wilson fermions  at pion mass
500~MeV  on a lattice of spatial length 1.8~fm at $\beta=6.0$. Left using $J_N$; Right using
$J_N^\prime$.}
\label{fig:nucleon}
\vspace*{-0.5cm}
\end{figure}

As a first new application of our method we examine the excited states of the nucleon using
local correlators that are produced easily in lattice simulations.
We use two interpolating fields:
\be
J_N(x)=\epsilon^{abc} (u_a C\gamma_5 d_b^T)u_c \quad, \hspace*{1cm}  
J^\prime_N(x)=\epsilon^{abc} (u_a^T C d_b)\gamma_5 u_c \quad.
\label{interpolating fields}
\ee
We compute the two-point correlators  with $J_N$ using two degenerate flavors of dynamical twisted
mass fermions (TMF)~\cite{ETMC} as well as  using
 two degenerate flavors of  dynamical Wilson fermions~\cite{Urbach}.
We apply our method taking $L=2$ and the resulting
spectrum for the positive and negative parity channels is shown in Fig.~\ref{fig:nucleon spectrum}.
The two lowest states can be clearly identified in both channels.
In addition, in the case of Wilson fermions, we compute the two-point
correlators  using  $J_N^\prime$.
It has been conjectured~\cite{Leinweber} 
that $J^\prime_N$ has a large overlap with the
Roper whereas its overlap with $J_N$ is small. We compare the
 distribution of amplitudes and masses in the positive parity channel 
in Fig.~\ref{fig:nucleon}  from correlators
 using these two interpolating fields 
for $\kappa=0.1580$ ($m_\pi\sim 500$~MeV).
One clearly identifies the first excited state from these low quality data.
In addition, 
we observe that  the state of lowest mass that is present in the mass spectrum of
the correlator computed with $J_N$ is absent when using $J^\prime_N$.
Instead the correlator with $J_N^\prime$ has a lowest state that does not show up
when using $J_N$.  The conjecture is that this state is the Roper.

\begin{figure}[ht]
\vspace*{-0.3cm}
\begin{minipage}{7cm}
\epsfxsize=6.5truecm
\epsfysize=7truecm
\ifpdf
{\mbox{\includegraphics[height=7.cm,width=6.5cm]{Nmass_vs_mpi}}}
\else
{\mbox{\epsfbox{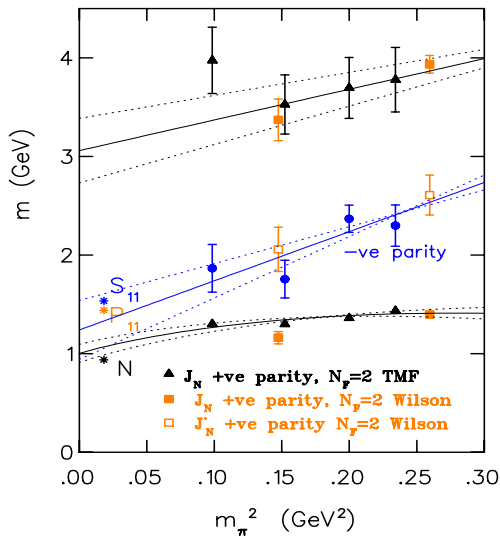}}}
\fi
\vspace*{-0.3cm}
\caption{Mass in the +ve and -ve parity channels of the nucleon  as function of $m_\pi^2$.}
\label{fig:Nmass}
\end{minipage}\hspace*{0.2cm}
\begin{minipage}{7.7cm}
\vspace*{-0.7cm}
%As already seen in the mass spectrum results of Fig.~\ref{fig:nucleon}
The mass of the first excited positive parity state extracted 
from the correlator with $J_N$, 
is too high to be identified with the
Roper. However, the mass of the lowest state extracted
from the correlator  with
$J_N^\prime$ is in the right mass range. It is
very close to  the mass of the negative
parity state and a linear extrapolation yields a value consistent
with the mass of the $P_{11}$ state, albeit with large statistical error.
This corroborates the conjecture that this state is the Roper.
In contrast to their different behavior for the ground state,
  both $J_N$ and $J_N^\prime$ interpolating fields
yield the same mass for the first excited state in the positive 
parity channel as well as for the two lowest states
  in the negative parity channels. 
\end{minipage}
\vspace*{-0.3cm}
\end{figure}

In Fig.~\ref{fig:Nmass} we show the mass of the two
lowest states in the positive parity channel and the mass of the lowest state
in the negative parity channel 
 as we vary the 
pion mass. Results 
 within these two lattice formulations
% the latter shown here only for the positive parity,
are compatible.
 We extrapolate TMF results  
on the ground state in
the positive parity channel to the physical point using lowest order heavy
baryon chiral perturbation theory, whereas we use a linear extrapolation
in $m_\pi^2$ for the negative parity   and first excited 
 positive parity states. We show with the solid
lines the best fit and with the dashed lines the error
band. At the physical point, within the
estimated errors,  the ground state masses of
 both positive  and
negative parity channels are  consistent with experiment, shown with
the asterisks.

\vspace*{-0.1cm}

\section{Conclusions}\vspace*{-0.3cm}
A novel method  for identifying and extracting parameters from lattice simulation data is presented. 
Applying the method to the
 $\eta_c$-correlator, we reproduce the results of a previous analysis 
where the use of priors is found to be important for the identification
of the two excited states. Our method, with no prior input, 
clearly identifies the two excited states  
with improved accuracy. Applying the method to the local nucleon correlators 
we determine the mass of the ground  
and the first excited states in the positive and negative parity channels.
Our results are in agreement with more evolved 
 mass correlation matrix analyzes yielding a positive parity state that
can be identified with the Roper.

\end{document}